\documentclass[prl,showpacs,reprint]{revtex4-1}
\usepackage{graphicx}
\usepackage{amsmath}
\usepackage{amssymb}

\newcommand{\ket}[1]{\ensuremath{|{#1}\rangle}}

\newcommand{\two}{\ensuremath{\ket{2,-2}}}
\newcommand{\aat}{\ket{a}}
\newcommand{\bat}{\ket{b}}

\begin{document}

\title{Collinear Four-Wave Mixing of Two-Component Matter Waves}

\author{Daniel Pertot}
\email{dpertot@ic.sunysb.edu}
\author{Bryce Gadway}
\author{Dominik Schneble}

\affiliation{Department of Physics~and~Astronomy, Stony Brook
University, Stony Brook, New York 11794-3800, USA}

\date{\today}

\begin{abstract}
We demonstrate atomic four-wave mixing of two-component matter waves in a
collinear geometry. Starting from a single-species Bose--Einstein condensate,
seed and pump modes are prepared through microwave state transfer and
state-selective Kapitza--Dirac diffraction. Four-wave mixing then populates the
initially empty output modes. Simulations based on a coupled-mode expansion of
the Gross--Pitaevskii equation are in very good agreement with the experimental
data. We show that four-wave mixing can play an important role in studies of
bosonic mixtures in optical lattices. Moreover our system should be of interest
in the context of quantum atom optics.
\end{abstract}

\pacs{03.75.Mn, 03.75.Gg, 67.85.Hj}

\maketitle

Four-wave mixing is a fundamental, well-studied concept in nonlinear
optics and spectroscopy~\cite{Boyd-2003-NonlinearOptics}. Its
matter-wave analogue, based on binary collisions in ultracold atomic
gases, was first demonstrated experimentally a decade
ago~\cite{Deng-1999-FourWaveMixingExpmt,Trippenbach-1998,*Trippenbach-2000-PRA},
establishing the field of nonlinear atom
optics~\cite{Meystre-2001-AtomOptics}. In four-wave mixing (FWM), two
waves form a grating from which a third wave diffracts, thus generating
a fourth wave. This process has been used for coherent matter-wave
amplification~\cite{VogelsXu-2002,*VogelsChin-2003}, and for the
generation of correlated atom pairs~\cite{Meystre-2001-AtomOptics,
VogelsXu-2002,*VogelsChin-2003,Perrin-2007-CollidingHeBECs,Dall-2009-HopeHeFWM}.
Energy and momentum conservation require the magnitudes of all atomic
momenta in the center-of-mass frame to be equal which, for atoms in a
single internal state, necessitates a two-dimensional
geometry~\cite{Trippenbach-1998,*Trippenbach-2000-PRA}. By modifying the
dispersion relation with an optical lattice, nondegenerate FWM of a
single species becomes possible also in one
dimension~\cite{Hillingsoe-2005,*Gemelke-2005-ParamAmplPeriodTranslLattices,%
*Campbell-2006-ParametricAmplification}.

Despite considerable theoretical work on atomic FWM
with more than one internal state~\cite{Goldstein-1998-CollinearFWM,%
PuMeystre-2000-MacroscopicAtomicEPR,Duan-2000-SqueezeEntangleAtomicBeams,Burke-2004-FWMwithManySpinStates},
experiments have only very recently started to explore possible
mechanisms for such FWM~\cite{Dall-2009-HopeHeFWM,Klempt-2009}. The
additional internal degree of freedom allows for degenerate FWM to occur
in one dimension, with pairs of waves in different internal states
sharing the same momentum mode, opening up possibilities to generate
nonclassical matter-wave states, e.g.\ with macroscopic spin
entanglement~\cite{PuMeystre-2000-MacroscopicAtomicEPR,*Duan-2000-SqueezeEntangleAtomicBeams}.
In this Letter, we demonstrate free-space collinear atomic FWM involving
two internal states with distinct, macroscopically populated momentum
modes.

Apart from the relevance for quantum atom optics, another important
context arises in experimental studies of bosonic mixtures in optical
lattices~\cite{Catani-2008-KRbBoseBoseSFMI,Weld-2009,McKayDeMarco-2009}.
These systems are of high interest not only in connection with
applications to quantum
magnetism~\cite{Kuklov-2003-SuperCounterFlowOL,*Altman-2003-TwoCompBHM,*Isacsson-2005-TwoCompBHM},
but also for studies of decoherence
mechanisms~\cite{Recati-2005-SpinBoson,*Orth-2009-SpinBoson}, and for
lattice thermometry~\cite{Weld-2009,McKayDeMarco-2009}. Most experiments
with ultracold atoms in optical lattices to date rely on time-of-flight
information. In particular, a sudden release from the lattice projects
the band populations onto plane-wave
states~\cite{Pedri-2001-BECOLExpansion}. We find that for a homonuclear
mixture of interacting superfluids, FWM processes can alter the expected
momentum-space distributions, masking or even mimicking \emph{in situ}
interaction effects.

\begin{figure}[!b]
  \includegraphics[width=7.7cm]{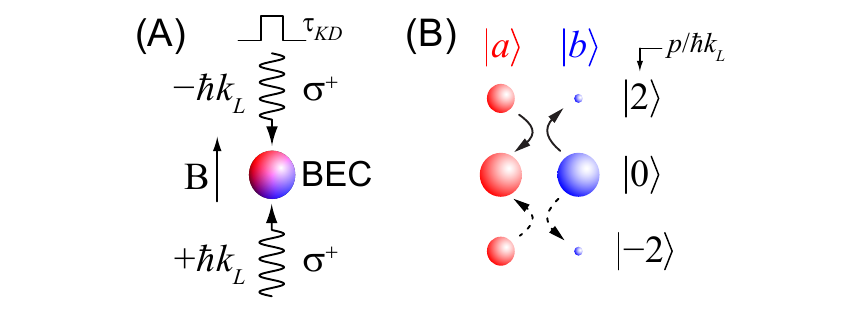}
  \caption{Experimental scheme. (A) State-selective
Kapitza--Dirac diffraction of a two-component Bose--Einstein condensate.
(B) Four-wave mixing (solid arrows) with pump modes $\ket{b, 0}$, $\ket{a, 2}$ and
seed mode $\ket{a, 0}$ transfers $\bat$ atoms to the output mode
$\ket{b, 2}$. Because of the symmetry of the problem, the process also
occurs for the modes $\ket{a, -2}$ and $\ket{b, -2}$ (dashed arrows).}
  \label{fig:scheme}
\end{figure}

In order to induce collinear two-component FWM, we apply a
state-selective optical lattice pulse to a Bose--Einstein condensate
containing atoms in two internal states $\aat$ and $\bat$, as
illustrated in Fig.~\ref{fig:scheme}~(A). The pulse induces
Kapitza--Dirac (KD)
diffraction~\cite{Gould-1986-AtomicKD,*Ovchinnikov-1999,Gadway-2009-KDbeyondRamanNath}
producing recoiling $\aat$ atoms in both positive and negative momentum
modes $\ket{\pm2}\equiv \ket{\pm2\hbar k_L}$ where $k_L=2\pi/\lambda_L$,
while the $\bat$ atoms remain unaffected. Subsequently, as illustrated
in Fig.~\ref{fig:scheme}~(B), the $\bat$ atoms Bragg diffract from the
density modulation formed by the interference of the recoiling $\aat$
atoms $\ket{a,2}\equiv\aat \otimes\ket{2}$ with those at rest,
$\ket{a,0}$. Because of momentum exchange collisions, the recoiling
$\aat$ atoms are coherently transferred back into $\ket{0}$, as
recoiling atoms $\ket{b,2}$ are produced. This process is formally not
distinguishable from coherent (pseudo-) spin exchange. Our system might
thus pose an interesting alternative to spinor condensates for the
creation of nonclassical
states~\cite{PuMeystre-2000-MacroscopicAtomicEPR,Duan-2000-SqueezeEntangleAtomicBeams,Klempt-2009}.
We note that due to the symmetry of the KD pulse, another, independent
``copy'' of the FWM process occurs on the negative momentum side. For
quantum atom optics purposes, this can easily be avoided by using a
state-selective Bragg pulse instead, which also allows for extended
control of the initial mode populations. In the present work, however,
we are content with applying a KD pulse, mainly out of technical
convenience.

Our experimental setup has been described in detail in
Ref.~\cite{Pertot-2009-MachinePaper}. In a crossed-beam optical dipole
trap at 1064~nm wavelength, we produce nearly pure
$^{87}$Rb~Bose--Einstein condensates in the $\aat \equiv
\ket{F=1,m_F=-1}$ hyperfine state typically containing about
$1.6\times10^5$ atoms. The trap is approximately isotropic with a mean
trap frequency around 50~Hz and an alignment-dependent vertical
frequency $\omega_z/2\pi$ between 40 and 50~Hz. Immediately after a
variable fraction of the condensate is transferred into the state $\bat
\equiv \two$ via a microwave Landau--Zener
sweep~\cite{Mewes-1997-LandauZener}, a state-selective lattice
beam~\cite{DeutschJessen-1998-QuantStateControlOptLat,*Jaksch-1999-ColdCollisions}
at $\lambda_L = 785.1$~nm is pulsed on along the vertical $z$ direction
($1/e^2$ radius 230~$\mu$m) for a time $\tau_{KD}$. The polarization
($\sigma^+$) is chosen such that only the $\aat$ atoms feel the optical
lattice potential formed through retro-reflection of the beam. A
magnetic field ($\sim$0.4~G) along the beam axis defines the
quantization axis. After release from the trap and a few milliseconds of
free evolution, during which the FWM occurs, a magnetic field gradient
(Stern--Gerlach pulse) spatially separates the two hyperfine states
along the horizontal $x$ axis for detection. The atoms are imaged after
a total time of flight of 15~ms via near-resonant absorption imaging by
a 100~$\mu$s long pulse of $F=2\rightarrow F'=3$ imaging light, combined
with $F=1\rightarrow F'=2$ repumping light, which ensures equal
detection efficiencies for both hyperfine states.

\begin{figure}[!b]
  \includegraphics[width=8cm]{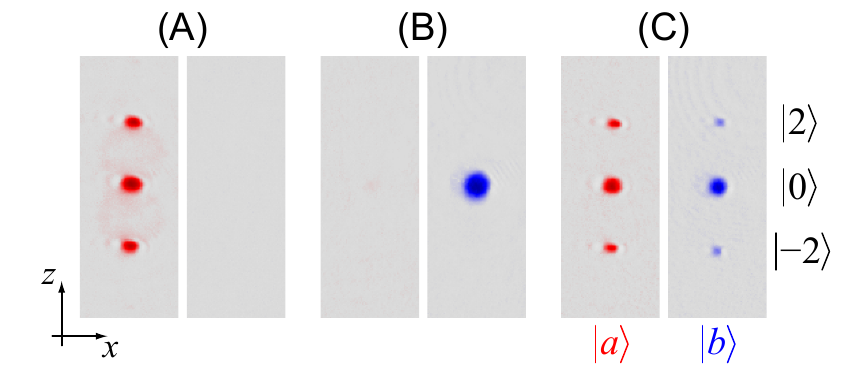}
  \caption{Typical absorption images taken after application of the state-selective
  Kapitza--Dirac pulse ($\tau_{KD}=25~\mu$s, $V_a=6$~$E_R$), 15~ms time of flight and Stern--Gerlach
separation along $x$, for the case of (A) only $\aat$ atoms present ($f_a=1$), (B)
only $\bat$ atoms ($f_a=0$), and (C) equal populations of both components ($f_a=0.5$).}
  \label{fig:pics}
\end{figure}

In Fig.~\ref{fig:pics}, typical absorption images are shown for three
different fractions of $\aat$ atoms $f_{a}\equiv N_a/N$. The KD pulse
duration (25~$\mu$s) and the lattice depth $V_a$ for atoms of type
$\aat$ (6~$E_R$, where $E_R=\hbar^{2} k_L^2 /2 m$ is the recoil energy),
are chosen such that half of the $\aat$ population is diffracted into
$\ket{\pm2}$, while higher orders remain largely unpopulated. By
analyzing single-component diffraction
patterns~\cite{Gadway-2009-KDbeyondRamanNath}, we have determined the
lattice depths for each component, confirming that atoms of type $\bat$
experience $<5$\% of the lattice depth seen by the $\aat$ atoms. On
their own, the $\bat$ atoms therefore are not affected by the lattice
pulse, as shown in Fig.~\ref{fig:pics}~(B). However, when both
components are present, a significant fraction of $\bat$ atoms is
transferred into the $\ket{\pm 2}$ momentum modes
[Fig.~\ref{fig:pics}~(C)].

\begin{figure}[!t]
  \includegraphics[width=7.8cm]{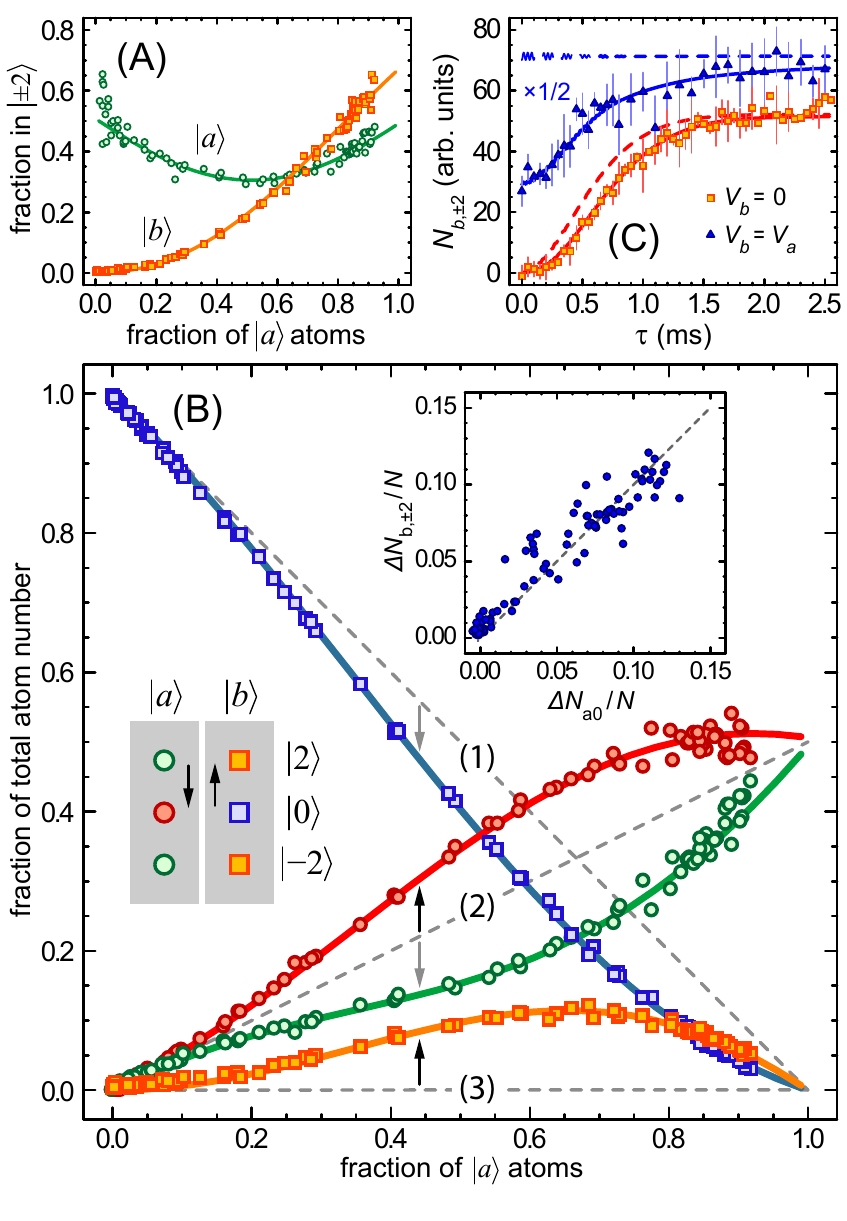}
  \caption{Mode populations after four-wave mixing.
(A) Fraction of atoms with momenta $\pm 2\hbar k_L$ in state $\aat$ (green circles)
and $\bat$ (orange squares).
(B) Populations $|c_{n\alpha}|^2=N_{n\alpha}/N$ of the modes
$\ket{\alpha, n}$, as indicated in the plaquette. The dashed lines (1), (2), and (3)
indicate the initial conditions before FWM,
$|b_0(0)|^2$, $|a_{0}(0)|^2=|a_{\pm2}(0)|^2$, and $|b_{\pm2}(0)|^2$ (where $\pm2$ indicates the combined populations). The arrows indicate
the temporal evolution of the populations. The solid lines represent the
predictions of the model ($V_a =5.6~E_R$, $\tau_{KD} = 25~\mu$s, $\omega_z = 2\pi\times 51$~Hz, $N=1.4\times 10^5$). The inset
shows the transferred $\bat$ population vs the transferred $\aat$ population,
where the dashed line represents a slope of unity.
(C) Growth of the population in $\ket{b, \pm2}$ following the Kapitza--Dirac
pulse for $V_b = 0$ (orange squares) and for $V_b = V_a$ (blue triangles, $\times 1/2$).
The FWM was interrupted after a variable time $\tau$ by blasting
away the $\aat$ atoms. Each data point is averaged
over 2--6 runs (here $\omega_z = 2\pi\times 41$~Hz, $f_a=0.5$). The dashed lines are
the predictions of the uncorrected model (including higher order FWM terms), whereas
the solid lines take into account
the loss of atoms during the blasting process (see text). The blast-loss
model was calibrated by fitting to the $V_b = V_a$ data.}
  \label{fig:data}
\end{figure}

We have measured the amount of diffracted atoms in each state as a
function of $f_a$. As shown in Fig.~\ref{fig:data}~(A), the fraction of
diffracted $\bat$ atoms $(N_{b,+2}+N_{b,-2})/N_b$ monotonically
increases from zero towards a maximum as $f_a$ is increased, consistent
with the picture that the grating formed by interference of the
$\ket{a,0}$ and $\ket{a,\pm2}$ modes, from which the atoms in
$\ket{b,0}$ diffract, gets deeper as the number of $\aat$ atoms grows.
The relative number of diffracted $\aat$ atoms has a pronounced minimum
near $f_a=0.5$, which can be interpreted as a ``backaction'' of the
$\bat$ atoms onto the $\aat$ grating.

To obtain a more quantitative understanding, we theoretically model our system
starting from the coupled Gross-Pitaevskii equations (GPE) for the order
parameters $\Phi_{\alpha}(\mathbf{r},t)$ of the two components $\alpha \in
\{a,b\}$
\begin{equation}\label{eq:GPE3d}\begin{split}
    i\hbar\,\partial_{t}\Phi_{\alpha}& = \left(\! -\frac{\hbar^{2}}{2m}\nabla^2 + V^{\mathrm{tot}}_{\alpha} + \!\!\sum_{\beta \in \{a,b\}}\!\! g_{\alpha\beta}\, |\Phi_{\beta}|^{2} \right)\Phi_{\alpha}, \raisetag{12pt}
\end{split}\end{equation}
where $g_{\alpha\beta}=4\pi\hbar^2 a_{\alpha\beta}/m$, $m$ is the atomic
mass, and the intra and interspecies $s$-wave scattering lengths
$a_{aa}$, $a_{bb}$, and $a_{ab}$ in units of $a_0$ are $100.4$, $99.0$,
and $99.0$, respectively~\cite{Kokkelmans-2010,*Verhaar-2009}. The
trapping and lattice potentials are given by $V^{\mathrm{tot}}_{\alpha}
= V_\mathrm{trap}(\mathbf{r},t) + V_{\alpha}(t)\,\sin^{2}\!{(k_L z)}$.
Similar to the slowly varying envelope approximation
(SVEA)~\cite{Trippenbach-1998,*Trippenbach-2000-PRA,Burke-2004-FWMwithManySpinStates},
we approximate the solution of Eq.~\eqref{eq:GPE3d} as an expansion in
terms of momentum modes, or wave packets, moving along $z$ with
multiples of the recoil velocity $v_R=\hbar k_L/m$
\begin{equation}\label{eq:ModeExp}
    \Phi_{\alpha}(\mathbf{r},t) = \sum_{n=-\infty}^{\infty} c_{n\alpha\!}(t)\, e^{i n k_{L\!} z}\, \Phi_0(\mathbf{r}-\hat{\mathbf{z}}\, n v_{R} t ,t)\, .
\end{equation}
We further assume that the wave packets $\Phi_0$ are of Thomas--Fermi
form and that they expand hydrodynamically after release from the
trap~\cite{Dalfovo-99}, which leads to a significant simplification
compared to a full SVEA simulation. On the time scales of interest,
phase-separation~\cite{Hall-1998-ComponentSeparation,*Hall-1998-RelativePhase,*Mertes-2007-HallCompSeparationReloaded}
can be neglected, and we have $\Phi_{\alpha}\propto\Phi_0$ for both
components just after the microwave
transfer~\cite{Hall-1998-ComponentSeparation,*Hall-1998-RelativePhase}.
Since the momentum spread of $\Phi_0$ is much less than $\hbar k_L$, the
modes in the expansion are quasi-orthogonal. After inserting the
ansatz~\eqref{eq:ModeExp} into Eq.~\eqref{eq:GPE3d}, we arrive at a
system of coupled equations for the amplitudes $a_n(t) \equiv c_{na}(t)$
\begin{displaymath}\begin{split}
    i\hbar\,\partial_{t} a_{n} &= E_R\, n^2 a_{n} + V_{a}(t) \left[\,\tfrac{1}{2}\, a_n - \tfrac{1}{4}\, (a_{n+2} + a_{n-2})\right]\\
                               &+\!\! \sum_{mm'n'}\!\!  (g_{aa}\, a_{m}^\ast a_{m'} + g_{ab}\, b_{m}^\ast b_{m'}) a_{n'}\, h_{n m m'n'}(t)
\end{split}\end{displaymath}
and similarly for the other component $b_n(t) \equiv c_{nb}(t)$. Here,
$h_{nmm'n'}(t) \propto \delta(n+m-m'-n')$ denotes overlap integrals that
include the effective temporal decay of the nonlinear interaction, as
the different wave packets separate, and as the density decreases during
the expansion. The terms responsible for FWM are of the form $b_m^\ast
b_n a_m$ (and $a_m^\ast a_n b_m$ for the $\bat$ component) with $m\neq
n$. After adjacent modes ($|m-n|=2$), for which the overlap decays the
slowest, have completely separated, the populations remain frozen, since
only equal-momentum self and cross-phase modulation terms of the forms
$|a_n|^2 a_n$ and $|b_n|^2 a_n$ survive. With a typical Thomas--Fermi
radius $R_z\sim 10$~$\mu$m, we obtain a typical separation time
$t_\mathrm{sep} \approx 2 R_z / 2 v_R$ of 1.7~ms.

The full set of observed populations $|a_n|^2$ and $|b_n|^2$ after FWM is
plotted in Fig.~\ref{fig:data}~(B), along with predictions of our model
obtained with parameters according to the experimental ones, leaving only
the total atom number $N$ as a fit parameter. The overall agreement
between data and theory is remarkable. The maximum FWM yield occurs near
$f_a = 2/3$ where the initial populations of the pump and seed modes are
equal, maximizing the FWM term $a_m^\ast a_n b_m$ at
$t=0$~\cite{Trippenbach-1998,*Trippenbach-2000-PRA}. The data also clearly
show the correlated growth of $|b_{\pm2}|^2$ and $|a_0|^2$, along with a
corresponding depletion of the pump modes $|b_0|^2$ and $|a_{\pm2}|^2$, as
detailed in the inset.

We note that since the FWM yield is proportional to the interspecies
scattering length as well as to the overlap $\int\!d\mathbf{r}\,
|\Phi_a|^2 |\Phi_b|^2$ of the two components, it can serve as a
sensitive probe for both quantities. As a practical example, we use
two-component FWM as a clear ``single-shot'' diagnostic for the
optimization of component overlap. By carefully canceling magnetic field
gradients, we are able to sustain overlap, i.e.\ FWM yield, for up to
2~s after the microwave transfer.

To further confirm the coherence of the observed two-component FWM as
implied by our model, we directly map out the time evolution of the
output mode population $|b_{\pm2}|^2$ by interrupting the FWM process
after a variable time through the selective removal of $\aat$ atoms with
a 50~$\mu$s long ``blast'' pulse of repumping light. As shown in
Fig.~\ref{fig:data}~(C), the atom number in the $\ket{b,\pm2}$ modes
smoothly grows from zero to a maximum value reached around the expected
separation time. The nonlinear, initially quadratic growth is indicative
of a coherent process~\cite{VogelsXu-2002,Trippenbach-2000-PRA} (other
signs would be an overshoot and oscillations, which however would
require higher densities or longer overlap). To exclude the possibility
that the observed growth is merely an artifact caused by
density-dependent losses of $\bat$ atoms accompanying the blast (due to
collisions with $\aat$ atoms), we repeat the experiment with the
polarization of the lattice beam chosen such that both components
experience the same lattice depth of about 6~$E_R$. In this case, we
expect the $\ket{b,\pm2}$ modes to be populated immediately after the KD
pulse, as indeed can be seen in Fig.~\ref{fig:data}~(C). Further, no FWM
is expected to occur for $V_b=V_a$, as the internal and external state
dynamics are decoupled. By comparing the observed time evolution for
this reference case with the expected one, we can calibrate our model
for the blast-induced losses, which assumes a relative loss of $\bat$
atoms proportional to the density of $\aat$ atoms in the overlap region.
With this correction, the theoretical time evolution for $V_b = 0$
matches the experimental data very well.

\begin{figure}[!t]
  \includegraphics[width=7.0cm]{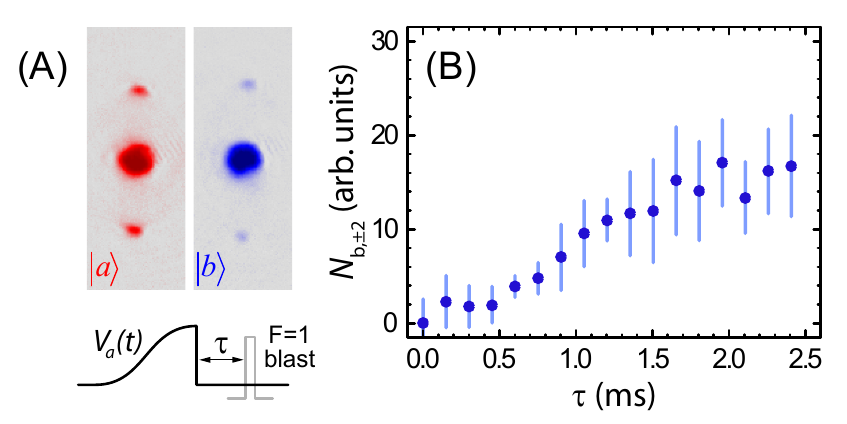}
  \caption{Four-wave mixing effects for an adiabatically ramped-up optical lattice.
  (A) An $\aat$-selective lattice is ramped up to a depth of 6.0~$E_R$ within 100~ms onto
  an equal mixture ($f_a=0.5$).
  After release and 17~ms time of flight, $\bat$ atoms appear in $\ket{\pm2}$.
  (B) Growth of population in $\ket{b,\pm2}$ as determined by blasting away the $\aat$ atoms
  after an evolution time $\tau$. Each data point is averaged over 6~runs.
  }
  \label{fig:4}
\end{figure}

So far, we have discussed controlled FWM after application of a short
optical pulse to induce diffraction. Now, we turn to the question
whether FWM is also relevant for adiabatically ramped-up,
state-selective optical lattices. For such a system, interspecies
interactions can be expected to give rise to diffraction effects
qualitatively similar to those due to FWM. The density profile of the
$\aat$ component gets spatially modulated by the optical lattice, thus
forming an ``atomic lattice'' that, in turn, should modulate the density
of the $\bat$ component, leading to diffraction peaks at $\pm2 \hbar
k_L$ immediately after release. However, we find that, at least as long
as both components are in the superfluid state, FWM is by far the
dominant mechanism for the emergence of recoiling $\bat$ atoms, caused
by the projection of the $\aat$ component into plane-wave momentum modes
after release. We note that the mismatch between the dispersion
relations for $\aat$ and $\bat$ atoms suppresses FWM while the lattice
is on. For $f_a=0.5$ and $V_a = 6~E_R$ ($V_b =0$), as shown in
Fig.~\ref{fig:4}~(A), we measure a relative population of up to $1.5~\%$
in each of the $\ket{b,\pm2}$ states. Assuming this to be caused by a
density modulation would require an atomic lattice modulation depth of
$2~E_R$, more than the chemical potential of the condensate in the
lattice. A blast measurement as discussed above shows that the
population in the observed peaks slowly grows only after release from
the lattice [Fig.~\ref{fig:4}~(B)], indicating that the peaks are indeed
caused by FWM.

We have performed analogous experiments for different final lattice
depths $V_a$, and with additional, state-independent lattices along the
$x$ and $y$ directions. These results will be presented in detail in a
future publication. In brief, we observe similar FWM effects along the
state-dependent axis (cf.~also~\cite{McKayDeMarco-2009}, Fig.~8);
however, we find that no FWM peaks are produced when the $\aat$~atoms
are in the Mott regime. This is consistent with the notion that FWM as
described relies on the existence of a well-defined macroscopic phase
and thus bears the potential to be used as a sensitive probe of phase
coherence.

To summarize, we have demonstrated collinear four-wave mixing in a
two-component mixture of bosonic atoms, and find excellent agreement with a
simple theoretical model. Our work is of relevance both in the context of
quantum atom optics, and for experimental studies of bosonic mixtures in
optical lattices.

\begin{acknowledgments}
We thank R.~Reimann for experimental contributions and T.~Bergeman for
careful reading of the manuscript. This work was supported by NSF
(PHY-0855643), ONR (DURIP), the Research Foundation of SUNY, and through
the GAANN program (B.G.) of the US ED.
\end{acknowledgments}

%


\begin{thebibliography}{10}%
\makeatletter
\providecommand \@ifxundefined [1]{%
 \ifx #1\undefined \expandafter \@firstoftwo
 \else \expandafter \@secondoftwo
\fi
}%
\providecommand \@ifnum [1]{%
 \ifnum #1\expandafter \@firstoftwo
 \else \expandafter \@secondoftwo
\fi
}%
\providecommand \enquote [1]{``#1''}%
\providecommand \bibnamefont  [1]{#1}%
\providecommand \bibfnamefont [1]{#1}%
\providecommand \citenamefont [1]{#1}%
\providecommand\href[0]{\@sanitize\@href}%
\providecommand\@href[1]{\endgroup\@@startlink{#1}\endgroup\@@href}%
\providecommand\@@href[1]{#1\@@endlink}%
\providecommand \@sanitize [0]{\begingroup\catcode`\&12\catcode`\#12\relax}%
\@ifxundefined \pdfoutput {\@firstoftwo}{%
 \@ifnum{\z@=\pdfoutput}{\@firstoftwo}{\@secondoftwo}%
}{%
 \providecommand\@@startlink[1]{\leavevmode}%
 \providecommand\@@endlink[0]{}%
}{%
 \providecommand\@@startlink[1]{%
  \leavevmode
  \pdfstartlink
   attr{/Border[0 0 1 ]/H/I/C[0 1 1]}%
   user{/Subtype/Link/A<</Type/Action/S/URI/URI(#1)>>}%
  \relax
 }%
 \providecommand\@@endlink[0]{\pdfendlink}%
}%
\providecommand \url  [0]{\begingroup\@sanitize \@url }%
\providecommand \@url [1]{\endgroup\@href {#1}{\urlprefix}}%
\providecommand \urlprefix [0]{URL }%
\providecommand \Eprint[0]{\href }%
\@ifxundefined \urlstyle {%
  \providecommand \doi [1]{doi:\discretionary{}{}{}#1}%
}{%
  \providecommand \doi [0]{doi:\discretionary{}{}{}\begingroup
  \urlstyle{rm}\Url }%
}%
\providecommand \doibase [0]{http://dx.doi.org/}%
\providecommand \Doi[1]{\href{\doibase#1}}%
\providecommand \bibAnnote [3]{%
  \BibitemShut{#1}%
  \begin{quotation}\noindent
    \textsc{Key:}\ #2\\\textsc{Annotation:}\ #3%
  \end{quotation}%
}%
\providecommand \bibAnnoteFile [2]{%
  \IfFileExists{#2}{\bibAnnote {#1} {#2} {\input{#2}}}{}%
}%
\providecommand \typeout [0]{\immediate \write \m@ne }%
\providecommand \selectlanguage [0]{\@gobble}%
\providecommand \bibinfo [0]{\@secondoftwo}%
\providecommand \bibfield [0]{\@secondoftwo}%
\providecommand \translation [1]{[#1]}%
\providecommand \BibitemOpen[0]{}%
\providecommand \bibitemStop [0]{}%
\providecommand \bibitemNoStop [0]{.\EOS\space}%
\providecommand \EOS [0]{\spacefactor3000\relax}%
\providecommand \BibitemShut [1]{\csname bibitem#1\endcsname}%
\bibitem{Boyd-2003-NonlinearOptics}%
  \BibitemOpen
  \bibfield{author}{%
  \bibinfo {author} {\bibfnamefont{R.~W.}\ \bibnamefont{Boyd}},\ }%
  \emph{\bibinfo {title} {Nonlinear Optics}}\ (\bibinfo {publisher} {Academic
  Press, San Diego},\ \bibinfo {year} {2003})%
  \bibAnnoteFile{NoStop}{Boyd-2003-NonlinearOptics}%
\bibitem{Deng-1999-FourWaveMixingExpmt}%
  \BibitemOpen
  \bibfield{author}{%
  \bibinfo {author} {\bibfnamefont{L.}~\bibnamefont{Deng}} \emph{et~al.},\ }%
  \bibfield{journal}{%
  \bibinfo {journal} {Nature}\ }%
  \textbf{\bibinfo {volume} {398}},\ \bibinfo {pages} {218} (\bibinfo {year}
  {1999})%
  \bibAnnoteFile{NoStop}{Deng-1999-FourWaveMixingExpmt}%
\bibitem{Trippenbach-1998}%
  \BibitemOpen
  \bibfield{author}{%
  \bibinfo {author} {\bibfnamefont{M.}~\bibnamefont{Trippenbach}}, \bibinfo
  {author} {\bibfnamefont{Y.~B.}\ \bibnamefont{Band}},\ and\ \bibinfo {author}
  {\bibfnamefont{P.~S.}\ \bibnamefont{Julienne}},\ }%
  \bibfield{journal}{%
  \bibinfo {journal} {Opt. Express}\ }%
  \textbf{\bibinfo {volume} {3}},\ \bibinfo {pages} {530} (\bibinfo {year}
  {1998})%
  \bibAnnoteFile{NoStop}{Trippenbach-1998}%
\bibitem{Trippenbach-2000-PRA}%
  \BibitemOpen
  \bibfield{author}{%
  \bibinfo {author} {\bibfnamefont{M.}~\bibnamefont{Trippenbach}}, \bibinfo
  {author} {\bibfnamefont{Y.~B.}\ \bibnamefont{Band}},\ and\ \bibinfo {author}
  {\bibfnamefont{P.~S.}\ \bibnamefont{Julienne}},\ }%
  \bibfield{journal}{%
  \bibinfo {journal} {Phys. Rev. A}\ }%
  \textbf{\bibinfo {volume} {62}},\ \bibinfo {pages} {023608} (\bibinfo {year}
  {2000})%
  \bibAnnoteFile{NoStop}{Trippenbach-2000-PRA}%
\bibitem{Meystre-2001-AtomOptics}%
  \BibitemOpen
  \bibfield{author}{%
  \bibinfo {author} {\bibfnamefont{P.}~\bibnamefont{Meystre}},\ }%
  \emph{\bibinfo {title} {Atom Optics}}\ (\bibinfo {publisher} {Springer, New
  York},\ \bibinfo {year} {2001})%
  \bibAnnoteFile{NoStop}{Meystre-2001-AtomOptics}%
\bibitem{VogelsXu-2002}%
  \BibitemOpen
  \bibfield{author}{%
  \bibinfo {author} {\bibfnamefont{J.~M.}\ \bibnamefont{Vogels}}, \bibinfo
  {author} {\bibfnamefont{K.}~\bibnamefont{Xu}},\ and\ \bibinfo {author}
  {\bibfnamefont{W.}~\bibnamefont{Ketterle}},\ }%
  \bibfield{journal}{%
  \bibinfo {journal} {Phys. Rev. Lett.}\ }%
  \textbf{\bibinfo {volume} {89}},\ \bibinfo {pages} {020401} (\bibinfo {year}
  {2002})%
  \bibAnnoteFile{NoStop}{VogelsXu-2002}%
\bibitem{VogelsChin-2003}%
  \BibitemOpen
  \bibfield{author}{%
  \bibinfo {author} {\bibfnamefont{J.~M.}\ \bibnamefont{Vogels}}, \bibinfo
  {author} {\bibfnamefont{J.~K.}\ \bibnamefont{Chin}},\ and\ \bibinfo {author}
  {\bibfnamefont{W.}~\bibnamefont{Ketterle}},\ }%
  \bibfield{journal}{%
  \bibinfo {journal} {\emph{ibid.}}\ }%
  \textbf{\bibinfo {volume} {90}},\ \bibinfo {pages} {030403} (\bibinfo {year}
  {2003})%
  \bibAnnoteFile{NoStop}{VogelsChin-2003}%
\bibitem{Perrin-2007-CollidingHeBECs}%
  \BibitemOpen
  \bibfield{author}{%
  \bibinfo {author} {\bibfnamefont{A.}~\bibnamefont{Perrin}} \emph{et~al.},\ }%
  \bibfield{journal}{%
  \bibinfo {journal} {Phys. Rev. Lett.}\ }%
  \textbf{\bibinfo {volume} {99}},\ \bibinfo {pages} {150405} (\bibinfo {year}
  {2007})%
  \bibAnnoteFile{NoStop}{Perrin-2007-CollidingHeBECs}%
\bibitem{Dall-2009-HopeHeFWM}%
  \BibitemOpen
  \bibfield{author}{%
  \bibinfo {author} {\bibfnamefont{R.~G.}\ \bibnamefont{Dall}} \emph{et~al.},\
  }%
  \bibfield{journal}{%
  \bibinfo {journal} {Phys. Rev. A}\ }%
  \textbf{\bibinfo {volume} {79}},\ \bibinfo {pages} {011601(R)} (\bibinfo {year}
  {2009})%
  \bibAnnoteFile{NoStop}{Dall-2009-HopeHeFWM}%
\bibitem{Hillingsoe-2005}%
  \BibitemOpen
  \bibfield{author}{%
  \bibinfo {author} {\bibfnamefont{K.~M.}\ \bibnamefont{Hilligs{\o}e}}\ and\
  \bibinfo {author} {\bibfnamefont{K.}~\bibnamefont{M{\o}lmer}},\ }%
  \bibfield{journal}{%
  \bibinfo {journal} {Phys. Rev. A}\ }%
  \textbf{\bibinfo {volume} {71}},\ \bibinfo {pages} {041602(R)} (\bibinfo {year}
  {2005})%
  \bibAnnoteFile{NoStop}{Hillingsoe-2005}%
\bibitem{Gemelke-2005-ParamAmplPeriodTranslLattices}%
  \BibitemOpen
  \bibfield{author}{%
  \bibinfo {author} {\bibfnamefont{N.}~\bibnamefont{Gemelke}} \emph{et~al.},\
  }%
  \bibfield{journal}{%
  \bibinfo {journal} {Phys. Rev. Lett.}\ }%
  \textbf{\bibinfo {volume} {95}},\ \bibinfo {pages} {170404} (\bibinfo {year}
  {2005})%
  \bibAnnoteFile{NoStop}{Gemelke-2005-ParamAmplPeriodTranslLattices}%
\bibitem{Campbell-2006-ParametricAmplification}%
  \BibitemOpen
  \bibfield{author}{%
  \bibinfo {author} {\bibfnamefont{G.~K.}\ \bibnamefont{Campbell}}
  \emph{et~al.},\ }%
  \bibfield{journal}{%
  \bibinfo {journal} {\emph{ibid.}}\ }%
  \textbf{\bibinfo {volume} {96}},\ \bibinfo {pages} {020406} (\bibinfo {year}
  {2006})%
  \bibAnnoteFile{NoStop}{Campbell-2006-ParametricAmplification}%
\bibitem{Goldstein-1998-CollinearFWM}%
  \BibitemOpen
  \bibfield{author}{%
  \bibinfo {author} {\bibfnamefont{E.~V.}\ \bibnamefont{Goldstein}}\ and\
  \bibinfo {author} {\bibfnamefont{P.}~\bibnamefont{Meystre}},\ }%
  \bibfield{journal}{%
  \bibinfo {journal} {Phys. Rev. A}\ }%
  \textbf{\bibinfo {volume} {59}},\ \bibinfo {pages} {1509} (\bibinfo {year}
  {1999})%
  \bibAnnoteFile{NoStop}{Goldstein-1998-CollinearFWM}%
\bibitem{PuMeystre-2000-MacroscopicAtomicEPR}%
  \BibitemOpen
  \bibfield{author}{%
  \bibinfo {author} {\bibfnamefont{H.}~\bibnamefont{Pu}}\ and\ \bibinfo
  {author} {\bibfnamefont{P.}~\bibnamefont{Meystre}},\ }%
  \bibfield{journal}{%
  \bibinfo {journal} {Phys. Rev. Lett.}\ }%
  \textbf{\bibinfo {volume} {85}},\ \bibinfo {pages} {3987} (\bibinfo {year}
  {2000})%
  \bibAnnoteFile{NoStop}{PuMeystre-2000-MacroscopicAtomicEPR}%
\bibitem{Duan-2000-SqueezeEntangleAtomicBeams}%
  \BibitemOpen
  \bibfield{author}{%
  \bibinfo {author} {\bibfnamefont{L.-M.}\ \bibnamefont{Duan}} \emph{et~al.},\
  }%
  \bibfield{journal}{%
  \bibinfo {journal} {\emph{ibid.}}\ }%
  \textbf{\bibinfo {volume} {85}},\ \bibinfo {pages} {3991} (\bibinfo {year}
  {2000})%
  \bibAnnoteFile{NoStop}{Duan-2000-SqueezeEntangleAtomicBeams}%
\bibitem{Burke-2004-FWMwithManySpinStates}%
  \BibitemOpen
  \bibfield{author}{%
  \bibinfo {author} {\bibfnamefont{J.~P.}\ \bibnamefont{Burke}} \emph{et~al.},\
  }%
  \bibfield{journal}{%
  \bibinfo {journal} {Phys. Rev. A}\ }%
  \textbf{\bibinfo {volume} {70}},\ \bibinfo {pages} {033606} (\bibinfo {year}
  {2004})%
  \bibAnnoteFile{NoStop}{Burke-2004-FWMwithManySpinStates}%
\bibitem{Klempt-2009}%
  \BibitemOpen
  \bibfield{author}{%
  \bibinfo {author} {\bibfnamefont{C.}~\bibnamefont{Klempt}} \emph{et~al.},\ }%
  \bibfield{journal}{%
  \bibinfo {journal} {Phys. Rev. Lett.}\ }%
  \textbf{\bibinfo {volume} {103}},\ \bibinfo {pages} {195302} (\bibinfo {year}
  {2009})%
  \bibAnnoteFile{NoStop}{Klempt-2009}%
\bibitem{Catani-2008-KRbBoseBoseSFMI}%
  \BibitemOpen
  \bibfield{author}{%
  \bibinfo {author} {\bibfnamefont{J.}~\bibnamefont{Catani}} \emph{et~al.},\ }%
  \bibfield{journal}{%
  \bibinfo {journal} {Phys. Rev. A}\ }%
  \textbf{\bibinfo {volume} {77}},\ \bibinfo {pages} {011603(R)} (\bibinfo {year}
  {2008})%
  \bibAnnoteFile{NoStop}{Catani-2008-KRbBoseBoseSFMI}%
\bibitem{Weld-2009}%
  \BibitemOpen
  \bibfield{author}{%
  \bibinfo {author} {\bibfnamefont{D.~M.}\ \bibnamefont{Weld}} \emph{et~al.},\
  }%
  \bibfield{journal}{%
  \bibinfo {journal} {Phys. Rev. Lett.}\ }%
  \textbf{\bibinfo {volume} {103}},\ \bibinfo {pages} {245301} (\bibinfo {year}
  {2009})%
  \bibAnnoteFile{NoStop}{Weld-2009}%
\bibitem{McKayDeMarco-2009}%
  \BibitemOpen
  \bibfield{author}{%
  \bibinfo {author} {\bibfnamefont{D.}~\bibnamefont{McKay}}\ and\ \bibinfo
  {author} {\bibfnamefont{B.}~\bibnamefont{DeMarco}},\ } %
  \Eprint{http://arxiv.org/abs/arXiv:0911.4143v1}{arXiv:0911.4143v1}%
 \bibinfo {note} { [New J. Phys. 12, 055013 (2010) (published)]}%
  \bibAnnoteFile{NoStop}{McKayDeMarco-2009}%
\bibitem{Kuklov-2003-SuperCounterFlowOL}%
  \BibitemOpen
  \bibfield{author}{%
  \bibinfo {author} {\bibfnamefont{A.~B.}\ \bibnamefont{Kuklov}}\ and\ \bibinfo
  {author} {\bibfnamefont{B.~V.}\ \bibnamefont{Svistunov}},\ }%
  \bibfield{journal}{%
  \bibinfo {journal} {Phys. Rev. Lett.}\ }%
  \textbf{\bibinfo {volume} {90}},\ \bibinfo {pages} {100401} (\bibinfo {year}
  {2003})%
  \bibAnnoteFile{NoStop}{Kuklov-2003-SuperCounterFlowOL}%
\bibitem{Altman-2003-TwoCompBHM}%
  \BibitemOpen
  \bibfield{author}{%
  \bibinfo {author} {\bibfnamefont{E.}~\bibnamefont{Altman}} \emph{et~al.},\ }%
  \bibfield{journal}{%
  \bibinfo {journal} {New J. Phys.}\ }%
  \textbf{\bibinfo {volume} {5}},\ \bibinfo {pages} {113} (\bibinfo {year}
  {2003})%
  \bibAnnoteFile{NoStop}{Altman-2003-TwoCompBHM}%
\bibitem{Isacsson-2005-TwoCompBHM}%
  \BibitemOpen
  \bibfield{author}{%
  \bibinfo {author} {\bibfnamefont{A.}~\bibnamefont{Isacsson}} \emph{et~al.},\
  }%
  \bibfield{journal}{%
  \bibinfo {journal} {Phys. Rev. B}\ }%
  \textbf{\bibinfo {volume} {72}},\ \bibinfo {pages} {184507} (\bibinfo {year}
  {2005})%
  \bibAnnoteFile{NoStop}{Isacsson-2005-TwoCompBHM}%
\bibitem{Recati-2005-SpinBoson}%
  \BibitemOpen
  \bibfield{author}{%
  \bibinfo {author} {\bibfnamefont{A.}~\bibnamefont{Recati}} \emph{et~al.},\ }%
  \bibfield{journal}{%
  \bibinfo {journal} {Phys. Rev. Lett.}\ }%
  \textbf{\bibinfo {volume} {94}},\ \bibinfo {pages} {040404} (\bibinfo {year}
  {2005})%
  \bibAnnoteFile{NoStop}{Recati-2005-SpinBoson}%
\bibitem{Orth-2009-SpinBoson}%
  \BibitemOpen
  \bibfield{author}{%
  \bibinfo {author} {\bibfnamefont{P.~P.}\ \bibnamefont{Orth}}, \bibinfo
  {author} {\bibfnamefont{I.}~\bibnamefont{Stanic}},\ and\ \bibinfo {author}
  {\bibfnamefont{K.}\ \bibnamefont{Le~Hur}},\ }%
  \bibfield{journal}{%
  \bibinfo {journal} {Phys. Rev. A}\ }%
  \textbf{\bibinfo {volume} {77}},\ \bibinfo {pages} {051601(R)} (\bibinfo {year}
  {2008})%
  \bibAnnoteFile{NoStop}{Orth-2009-SpinBoson}%
\bibitem{Pedri-2001-BECOLExpansion}%
  \BibitemOpen
  \bibfield{author}{%
  \bibinfo {author} {\bibfnamefont{P.}~\bibnamefont{Pedri}} \emph{et~al.},\ }%
  \bibfield{journal}{%
  \bibinfo {journal} {Phys. Rev. Lett.}\ }%
  \textbf{\bibinfo {volume} {87}},\ \bibinfo {pages} {220401} (\bibinfo {year}
  {2001})%
  \bibAnnoteFile{NoStop}{Pedri-2001-BECOLExpansion}%
\bibitem{Gould-1986-AtomicKD}%
  \BibitemOpen
  \bibfield{author}{%
  \bibinfo {author} {\bibfnamefont{P.~L.}\ \bibnamefont{Gould}}, \bibinfo
  {author} {\bibfnamefont{G.~A.}\ \bibnamefont{Ruff}},\ and\ \bibinfo {author}
  {\bibfnamefont{D.~E.}\ \bibnamefont{Pritchard}},\ }%
  \bibfield{journal}{%
  \bibinfo {journal} {Phys. Rev. Lett.}\ }%
  \textbf{\bibinfo {volume} {56}},\ \bibinfo {pages} {827} (\bibinfo {year}
  {1986})%
  \bibAnnoteFile{NoStop}{Gould-1986-AtomicKD}%
\bibitem{Ovchinnikov-1999}%
  \BibitemOpen
  \bibfield{author}{%
  \bibinfo {author} {\bibfnamefont{Y.~B.}\ \bibnamefont{Ovchinnikov}}
  \emph{et~al.},\ }%
  \bibfield{journal}{%
  \bibinfo {journal} {\emph{ibid.}}\ }%
  \textbf{\bibinfo {volume} {83}},\ \bibinfo {pages} {284} (\bibinfo {year}
  {1999})%
  \bibAnnoteFile{NoStop}{Ovchinnikov-1999}%
\bibitem{Gadway-2009-KDbeyondRamanNath}%
  \BibitemOpen
  \bibfield{author}{%
  \bibinfo {author} {\bibfnamefont{B.}~\bibnamefont{Gadway}} \emph{et~al.},\ }%
  \bibfield{journal}{%
  \bibinfo {journal} {Opt. Express}\ }%
  \textbf{\bibinfo {volume} {17}},\ \bibinfo {pages} {19173} (\bibinfo {year}
  {2009})%
  \bibAnnoteFile{NoStop}{Gadway-2009-KDbeyondRamanNath}%
\bibitem{Pertot-2009-MachinePaper}%
  \BibitemOpen
  \bibfield{author}{%
  \bibinfo {author} {\bibfnamefont{D.}~\bibnamefont{Pertot}} \emph{et~al.},\ }%
  \bibfield{journal}{%
  \bibinfo {journal} {J. Phys. B}\ }%
  \textbf{\bibinfo {volume} {42}},\ \bibinfo {pages} {215305} (\bibinfo {year}
  {2009})%
  \bibAnnoteFile{NoStop}{Pertot-2009-MachinePaper}%
\bibitem{Mewes-1997-LandauZener}%
  \BibitemOpen
  \bibfield{author}{%
  \bibinfo {author} {\bibfnamefont{M.-O.}\ \bibnamefont{Mewes}} \emph{et~al.},\
  }%
  \bibfield{journal}{%
  \bibinfo {journal} {Phys. Rev. Lett.}\ }%
  \textbf{\bibinfo {volume} {78}},\ \bibinfo {pages} {582} (\bibinfo {year}
  {1997})%
  \bibAnnoteFile{NoStop}{Mewes-1997-LandauZener}%
\bibitem{DeutschJessen-1998-QuantStateControlOptLat}%
  \BibitemOpen
  \bibfield{author}{%
  \bibinfo {author} {\bibfnamefont{I.~H.}\ \bibnamefont{Deutsch}}\ and\
  \bibinfo {author} {\bibfnamefont{P.~S.}\ \bibnamefont{Jessen}},\ }%
  \bibfield{journal}{%
  \bibinfo {journal} {Phys. Rev. A}\ }%
  \textbf{\bibinfo {volume} {57}},\ \bibinfo {pages} {1972} (\bibinfo {year}
  {1998})%
  \bibAnnoteFile{NoStop}{DeutschJessen-1998-QuantStateControlOptLat}%
\bibitem{Jaksch-1999-ColdCollisions}%
  \BibitemOpen
  \bibfield{author}{%
  \bibinfo {author} {\bibfnamefont{D.}~\bibnamefont{Jaksch}} \emph{et~al.},\ }%
  \bibfield{journal}{%
  \bibinfo {journal} {Phys. Rev. Lett.}\ }%
  \textbf{\bibinfo {volume} {82}},\ \bibinfo {pages} {1975} (\bibinfo {year}
  {1999})%
  \bibAnnoteFile{NoStop}{Jaksch-1999-ColdCollisions}%
\bibitem{Kokkelmans-2010}%
  \BibitemOpen
  \bibfield{author}{%
  \bibinfo {author} {\bibfnamefont{S.~J. J. M.~F.}\ \bibnamefont{Kokkelmans}}\ }%
  \bibinfo {note} {(private communication)}%
  \bibAnnoteFile{NoStop}{Kokkelmans-2010}%
\bibitem{Verhaar-2009}%
  \BibitemOpen
  \bibfield{author}{%
  \bibinfo {author} {\bibfnamefont{B.~J.}\ \bibnamefont{Verhaar}}, \bibinfo
  {author} {\bibfnamefont{E.~G.~M.}\ \bibnamefont{van Kempen}},\ and\ \bibinfo
  {author} {\bibfnamefont{S.~J. J. M.~F.}\ \bibnamefont{Kokkelmans}},\ }%
  \bibfield{journal}{%
  \bibinfo {journal} {Phys. Rev. A}\ }%
  \textbf{\bibinfo {volume} {79}},\ \bibinfo {pages} {032711} (\bibinfo {year}
  {2009})%
  \bibAnnoteFile{NoStop}{Verhaar-2009}%
\bibitem{Dalfovo-99}%
  \BibitemOpen
  \bibfield{author}{%
  \bibinfo {author} {\bibfnamefont{F.}~\bibnamefont{Dalfovo}} \emph{et~al.},\
  }%
  \bibfield{journal}{%
  \bibinfo {journal} {Rev. Mod. Phys.}\ }%
  \textbf{\bibinfo {volume} {71}},\ \bibinfo {pages} {463} (\bibinfo {year}
  {1999})%
  \bibAnnoteFile{NoStop}{Dalfovo-99}%
\bibitem{Hall-1998-ComponentSeparation}%
  \BibitemOpen
  \bibfield{author}{%
  \bibinfo {author} {\bibfnamefont{D.~S.}\ \bibnamefont{Hall}} \emph{et~al.},\
  }%
  \bibfield{journal}{%
  \bibinfo {journal} {Phys. Rev. Lett.}\ }%
  \textbf{\bibinfo {volume} {81}},\ \bibinfo {pages} {1539} (\bibinfo {year}
  {1998})%
  \bibAnnoteFile{NoStop}{Hall-1998-ComponentSeparation}%
\bibitem{Hall-1998-RelativePhase}%
  \BibitemOpen
  \bibfield{author}{%
  \bibinfo {author} {\bibfnamefont{D.~S.}\ \bibnamefont{Hall}} \emph{et~al.},\
  }%
  \bibfield{journal}{%
  \bibinfo {journal} {Phys. Rev. Lett.}\ }%
  \textbf{\bibinfo {volume} {81}},\ \bibinfo {pages} {1543} (\bibinfo {year}
  {1998})%
  \bibAnnoteFile{NoStop}{Hall-1998-RelativePhase}%
\bibitem{Mertes-2007-HallCompSeparationReloaded}%
  \BibitemOpen
  \bibfield{author}{%
  \bibinfo {author} {\bibfnamefont{K.~M.}\ \bibnamefont{Mertes}}
  \emph{et~al.},\ }%
  \bibfield{journal}{%
  \bibinfo {journal} {\emph{ibid.}}\ }%
  \textbf{\bibinfo {volume} {99}},\ \bibinfo {pages} {190402} (\bibinfo {year}
  {2007})%
  \bibAnnoteFile{NoStop}{Mertes-2007-HallCompSeparationReloaded}%
\end{thebibliography}
\end{document}